\documentstyle[12pt]{article}
\def\zetZ{Z\!\!\!Z}
\def\theequation{\arabic{section}.\arabic{equation}}
\newcounter{subequation}[equation]
\makeatletter
\@addtoreset{equation}{section}
\expandafter\let\expandafter\reset@font\csname reset@font\endcsname
  {\endeqnarray\stepcounter{equation}}
\makeatother

\normalbaselines 
\begin{document}

\title{Translational Chern--Simons Action\\ and New  Planar Particle
Dynamics}

\author{J. Lukierski\\ \\
Institute for Theoretical Physics, University of Wroc\l aw, \\
 pl. Maxa Borna 9, 50-204 Wroc\l aw,
  Poland
  \\
  {email:lukier@ift.uni.wroc.pl}
\\ \\
P.C. Stichel\\ \\
An der Krebskuhle 21\\ D-33619 Bielefeld\\
{email:pstichel@gmx.de}
\\ \\
W.J. Zakrzewski
\\ \\
Department of Mathematical Sciences, University of Durham, \\
Durham DH1 3LE, UK
\\
{email:W.J.Zakrzewski@durham.ac.uk}}

\date{}
\maketitle
\thispagestyle{empty}
\begin{abstract}
We consider a nonstandard $D=2+1$ gravity  described by
a translational Chern--Simons action, and couple it to the
nonrelativistic point particles. 
We fix the asymptotic coordinate transformations in such a way that
the space part of the metric becomes asymptotically Euclidean.
The residual symmetries are (local in time) translations  and
rigid rotations.
The phase space Hamiltonian $H$
describing two-body interactions satisfies a nonlinear equation
$H={\cal H}( \vec{x},\vec{p};H)$ what implies, after quantization,
a nonstandard form of the Schr\"{o}dinger equation with
energy-dependent fractional angular momentum eigenvalues. 
Quantum solutions of the two-body problem are discussed. The bound
states with discrete energy levels correspond to a confined
classical motion (for the planar distance between two particles
$r\leq r_0$)  and the scattering states with continuous energy
correspond to classical motion for $r>r_0$.
\end{abstract}
\setcounter{page}{0}
\section{Introduction}

The aim of this paper is to study a new type of gravitational
point particle interactions in two space dimensions.   The classical and
quantum dynamics of point sources interacting with 
standard  (2+1)-dimensional gravity has been well known
since the papers by Staruszkiewicz, Deser, Jackiw and t`Hooft 
were published (see [1-4]).

In this paper we consider a new system of coupled
two-dimensional torsion fields with point particles, derived in the
following way:

i) We assume a nonstandard gravitational action described by
a translational Chern--Simons term [5-7]

\begin{equation}
\mbox{\sl S}_{T} 
= {1\over \lambda} \int d^{3}x \,
\varepsilon^{\mu\nu\rho} \, E^{\underline{\alpha}}_{\mu} \,
T^{\underline{\beta}}_{\nu\rho} \, 
\eta_{\underline{\alpha}\underline{\beta}},
\end{equation}
where $E^{\underline{\nu}}_{\mu} $ ($\overline{\nu}=0,1,2;
\mu=0,1,2$) describes (2+1)--dimensional dreibeins, and the
Abelian strength

\begin{equation}
T_{\mu\nu}^{\ \underline{\rho}} =
\partial_{\mu}
E^{\underline{\rho}}_{\nu} - 
\partial_{\nu}
E^{\underline{\rho}}_{\mu}
\end{equation}
provides the components of a (2+1)-dimensional  torsion 
tensor field. Note that a 
linear torsion Lagrangian taking the form of a translational
Chern-Simons term can be introduced only in (2+1) dimensions. 

Here we would like to recall that, following
Witten [5], the standard gravitational Einstein action in (2+1)
dimensions is described by another Chern--Simons theory

\begin{equation}
\mbox{\sl S}_{E} 
= {1\over \lambda} \int d^{3}x \,
\varepsilon^{\mu\nu\rho} \, E^{\underline{\alpha}}_{\mu} \,
\omega^{\underline{\beta}}_{\nu\rho} \, \eta_{\underline{\alpha}
\underline{\beta}},
\end{equation}
where $\omega^{\underline{\alpha}}_{\mu\nu} $
are the $O(2,1)$ spin connection components.

ii) In this paper we consider gravitational
interactions of 
 \underline{nonrelativistic} $D=2$ particles described by the
trajectory $\vec{x}(t)=(x_1 (t), x_2 (t))$, which are covariant
under the following set of nonrelativistic local transformations:

\begin{equation}
x^{\prime}_{l} = x^{\prime}_{l}(\vec{x}, t)\, ,
 \qquad t^{\prime} = t+a\, ,
\end{equation}
which imply that

\begin{equation}
\dot{x}^{\prime}_{i}= {\partial x^{\prime}_{i}\over
\partial x_{j}}  
\, \dot{x}^{\prime}_{j}
+{\partial x^{\prime}_{i}\over
\partial t}\, ,  \qquad{ dt^{\prime}\over dt} = 1 \, .
\end{equation}
The set of nonrelativistic reparametrization--invariant
velocities then takes the form $(\underline{a}, \underline{b}=1,2)$:

\begin{equation}
\xi^{\underline{a}} = 
E^{\underline{a}}_{\ i}\, \dot{x}_{i} +
E^{\underline{a}}_{\ 0} \, ,
\end{equation}
which can be obtained from the relativistic
reparametrization--invariant velocities

\begin{equation}
{\dot{x}}^{\mu} \to
\xi^{\underline{\mu}} 
= E^{\underline{\mu}}_{\ \nu} 
\, {\dot{x}}^{\nu}
=\left( \xi^{\underline{0}}, 
\xi^{\underline{a}} \right)\, ,
\end{equation}
by putting $x_0 = t$ and introducing a nonrelativistic gauge
fixing condition $E^{\underline{0}}_{\ 0}=1$, 
 $E^{\underline{0}}_{\ i}=0$. In  such a gauge we obtain the
nonrelativistic form of the action (1.1), invariant under (1.4)

\begin{equation}
\mbox{\sl S}^{\rm (NR)}_{\ T}= 
{1 \over \lambda} \int 
d^{3}x
 e^{\mu\nu\rho} 
  E^{\underline{a}}_{\mu}
   T^{\underline{a}}_{\nu\rho}
   =
{1 \over \lambda} \int dt
d^{2}x
\left( e^{\underline{a}}  B^{\underline{a}} -
\varepsilon_{jk} \, h^{\underline{q}}_{j} 
{\cal E}^{\underline{a}}_{k}  \right),
\end{equation}
where
\begin{equation}
{B}^{\underline{ a}} =
\varepsilon_{jk} \, \partial_{j} \, h^{\underline{a}}_{k}
\qquad
{\cal E}^{\underline{ a}}_{k} =
 \partial_{t} \, h^{\underline{a}}_{k}-
  \partial_{k} \, e^{\underline{a}}\, .
  \end{equation}
We denote $h^{\underline{a}}_{\kappa} = E^{ \underline{a}}_{\kappa}
$ and $e^{\underline{a}} = E^{ \underline{a}}_{0}
$.  The fields ${B}^{\underline{ a}}$ and ${\cal E}^{\underline{
a}}_{k}$ play the role of the magnetic and electric fields, respectively, 
with the internal $O(2)$ index $\underline{a}$ describing the $D=2$
nonrelativistic rotation group.

In Sect. 2 we discuss a coupled classical
 field + point particles system.
Firstly, we  present general considerations 
 for the $N$--body particle dynamics, and then we 
  consider  the two--body problem in
classical theory. 
 In this case 
we obtain, in terms of relative planar coordinates, 
a new type of dynamics which can be understood in
two equivalent ways:

$\bullet$ as a $D=2$ free particle 
 system with nonstandard,  symplectic
structure 

$\bullet$ as a modified $D=2$ free particle system with
an energy-dependent angular momentum $\overline{l}(E)$ and
a canonical symplectic structure.

We also show that the classical solutions split into two classes
for $\lambda \overline{l} < 0$

$\bullet$ for the planar particle distances 
$$ r < r_0 =  \left( { |\lambda \overline{l}(E)| \over 2
\pi} \right)^{1\over 2} $$
they describe bounded motion while 

$\bullet$ for $r>r_0$ they describe the  scattering.

In Sect. 3 we  introduce  the corresponding quantum mechanics,
  described by a 
 nonstandard Schr\"{o}dinger equation, and discuss the energy
eigenvalues and  wave eigenfunctions.

We find that the exact values of the energy levels for the confined
 motion ($r < r_0$)  can be
determined only numerically.

The final remarks, concerning possible physical interpretations
of our solutions are presented in the last, fourth Section.

It should be added that the present model is a
generalization of the $ (1+1)$--dimensional model, presented by one
of the present authors (PCS) in [8,9].

 \section{New Planar Particle Dynamics: Classical Theory}
  \subsection{${\bf N}$--Particle System}

 Let us
discuss, in some detail, a set of coupled equations
describing an interacting two-dimensional particle -- torsion 
field system, with
the torsion fields being given by the Abelian Chern--Simons action (1.3).
Introducing $N$ trajectories $\vec{x}_{\alpha}(t) =
\left\{ x^{i}_{\alpha}(t) \right\}$
 ($i=1,2; \alpha=1,\ldots N$) for $N$ particles and the notation
 \begin{equation}
 E^{\underline{a}}_{\mu; \alpha}(t) \equiv
 E^{\underline{a}}_{\mu}(\vec{x}_{\alpha}(t), t),
 \end{equation}
we find that the  action for $N$ particles in $D=2$ dimensions, in 
the first order formalism,
can be written as 
\begin{equation}
\mbox{\sl S}_{\rm part}^{\ (N)}= -
\sum\limits^{N}_{\alpha = 1} m_{\alpha}
\int dt \left[ {1\over 2} \xi^{\underline{a}}_{\alpha}
\, \xi^{\underline{a}}_{\alpha} -
\xi^{\underline{a}}_{\alpha}\left(
E^{\underline{a}}_{j; \alpha}\, \dot{x}^{j}_{\alpha} +
E^{\underline{a}}_{0, \alpha} \right)\right],
\end{equation}
thus providing the constraint formula
\begin{equation}
\xi^{\underline{a}}_{\alpha} = E^{\underline{a}}_{j; \alpha}
\dot{x}^{j} + E^{\underline{a}}_{0, \alpha}. 
\end{equation}


We can now derive the equations describing the $D=2$
coupled particle--torsion  field system, described by the action 
$S= \mbox{\sl S}^{\ (N)}_{\rm part}+
\mbox{\sl S}^{\ \rm (NR)}_{\ \rm T}$ (see (1.8) and (2.2)). They
 take the form:

\begin{equation}
\dot \xi_{\alpha}\sp{\underline{a}}\cdot E_{i, \alpha}\sp{\underline{a}}
\,+\,\xi_{\alpha}\sp{\underline{a}}\cdot T_{\mu i,\alpha}\sp{\underline{a}}
\dot x\sp{\mu,\alpha}\,=\,0,
\label{two} 
\end{equation}

\begin{equation}
T_{\mu \nu}\sp{\underline{a}}\,=
\partial_{\mu} 
E^{\underline{a}}_{\nu} 
-
\partial_{\nu} 
E^{\underline{a}}_{\mu}
= \,
\,-{\lambda\over 2}\,\epsilon_{\mu\nu\rho}
\sum\limits_{\beta} \xi_{\beta}\sp{\underline{a}}\,
\dot x\sp{\rho,\beta}\,\delta(\vec{x}-\vec{x}\sp{
\beta}).
\label{(2.7)}
\end{equation}
The second term in (\ref{two}), due to the antisymmetry of 
 the $\epsilon$--tensor in (\ref{(2.7)}), can be rewritten as

\begin{equation}
\xi_{\alpha}\sp{\underline{a}}\cdot T_{\mu i,\alpha}\sp{\underline{a}}
\dot x\sp{\mu,\alpha}\,=\,
-{\lambda\over 2}\epsilon_{\mu i\rho}\sum\limits_{\beta  \atop
\beta \ne \alpha}
\xi_{\beta}\sp{\underline{a}}\, 
\xi_{\alpha}\sp{\underline{a}}\,
\dot x\sp{\rho,\beta}
\dot x\sp{\mu,\alpha}\delta(\vec{x}_{\alpha}-
\vec{x}_{\beta}), \label{2.8}
\end{equation}
which is infinite for coinciding particle positions and vanishes otherwise. 
If we restrict our configuration space  only to
noncoinciding particle positions we have

\begin{equation}
\dot \xi_{\alpha}\sp{\underline{a}}
\cdot E_{\alpha;i}\sp{\underline{a}}\,=\,0,
\label{NNew} 
\end{equation}
which leads, for points in the configuration 
space where the metric is non-degenerate, to
\begin{equation}
\dot \xi_{\alpha}\sp{\underline{a}}\,=\,0.
\label{New} 
\end{equation}

Let us now consider the field equation (\ref{(2.7)}).
 Its general solution  can be written in the pure gauge
 form
 \begin{equation}
E^{\underline{a}}_{\mu}(\vec{x},t) = 
{\widetilde{E}}^{\underline{a}}_{\mu}(\vec{x},t)
+ \partial_{\mu} \Lambda^{\underline{a}}(\vec{x},t) = 
\partial_{\mu}\widetilde \Lambda^{\underline{a}}(\vec{x},t),
 \end{equation}
 where 
 \begin{equation}
{ \widetilde{E}}^{\underline{a}}_{\mu}(\vec{x},t)
 : =
-{\lambda \over 4\pi} \partial_{\mu}
 \sum\limits_{\alpha} \xi^{\underline{a}}_{\alpha}
\,\Phi(\vec{x}- \vec{x}_{\alpha})
\label{2.10nowe}
 \end{equation}
and
   
   - $\Lambda^{\underline{a}}$ is an $O(2)$--vector valued pair
of regular gauge functions,

- $\Phi(\vec{x})$ is a singular gauge function satisfying the
following equation (see e.g. [10]):
 \begin{equation}
\epsilon^{ij}   \partial_{i}
\partial_{j}
\Phi(\vec{x}) = 2\pi \, \delta(\vec{x}\, ).  
 \end{equation}
As a solution of  (2.11) we can
take
\begin{equation}
\Phi(\vec{x}\, ) = \mbox{arc}\,\mbox{tan} {x_{2}\over x_{1}}
\end{equation}
{\it i.e.}
\begin{equation}
\partial_{k}\Phi({x})  =- \varepsilon_{kl}\partial_{l}\ln
|\vec{x}\, |.  
\end{equation}
where (2.13) has to be regularized in such a way that it is well
defined everywhere and therefore vanishes for $\vec{x} \to 0$
(see e.g. [11]).

Let us note that the solutions for the fields  
$E^{\underline{a}}_{\mu}(\vec{x},t)$
with asymptotically nonvanishing gauge function $\Lambda^{a}$
are not solutions of the Hamilton's variational principle for
the action $S$. The bad asymptotic (as $r \to  \infty$)
 behaviour of 
$ E^{\underline{a}}_{\mu}(\vec{x},t)$ leads to the appearance of
nonvanishing surface integrals  
  and, in consequence, our 
$E^{\underline{a}}_{\mu}(\vec{x},t)$ do not minimize the
action. 
 In order to take into consideration the particular asymptotic
behaviour of $\Lambda^{a}$ we fix our gauge as follows:
\begin{equation}
\Lambda^{i}(\vec{x},t)= x^i - a^i (t)
\end{equation}
and modify the field Lagrangian by adding two surface
integrals\footnote {see [12] for such a procedure.}
\begin{eqnarray}
I_1 &= & {1\over \lambda} \int d^2x\, v^{\b{a}} \cdot
 \epsilon^{ij} \, \partial_i 
\,
{\widetilde{E}}^{\underline{a}}_{j}(\vec{x},t)
\\ \cr
I_2  &= & {1\over \lambda} \int d^2x\, \epsilon^{ij} \, \partial_i 
\,
{\widetilde{E}}^{j}_{0}
\end{eqnarray}
where we have defined
\begin{equation}
v ^{\underline{a}}(t) := \dot{a}^{\underline{a}} (t)\, .
\end{equation}

Notice that with the gauge (2.14) the space part of our metric
becomes asymptotically Euclidean. Due to the asymptotic behaviour
\begin{equation}
{\widetilde{E}}^{\underline{a}}_{\mu}(\vec{x},t) \to O(r^{-1}) \qquad
\hbox{as} \qquad r\to \infty
\end{equation}
the variations of the ${\widetilde{E}}_{\mu}^{\underline{a}}$ and of the
$v^{\underline{a}}$ are independent of each other i.e. the
 fields
${\widetilde{E}}^{\underline{a}}_{\mu}$ and $v^{\underline{a}}$ can be
introduced as new variables. The particle action (2.2) takes the form

\setcounter{equation}{0}
\renewcommand{\theequation}{\thesection.19\alph{equation}}
\begin{equation}
\mbox{\sl S}^{\rm (N)}_{\rm part} [x,\xi, E]
= \mbox{\sl S}^{\rm (N)}_{\rm part} [x,\xi, {\widetilde{E}}] +
\int dt \left[
\sum\limits_{\alpha} 
 \xi_{\alpha}^{\underline{a}}\,
\cdot \dot{x}^{\underline{a}}_{\alpha}
- v^{\underline{a}} \cdot
\sum\limits_{\alpha} 
 \xi_{\alpha}^{\underline{a}}
 \right].
 \end{equation}
 where as the modified field action $S_{\rm field}(\widetilde{E})$ 
  is given by $S^{(NR)}_{T}$ (eq. (1.8)) but taken
   as a function of the $\widetilde{E}$ now
   \begin{equation}
   \mbox{\sl S}_{\rm field} (\widetilde{E})  =  
      \mbox{\sl S}^{(NR)}_{T} (\widetilde{E}) 
      \end{equation}

\setcounter{equation}{19}
\renewcommand{\theequation}{\thesection.\arabic{equation}}

We see that in the action (2.19) we have separated the
variables describing the ``bulk'' $(\widetilde{E}^{a}_{\mu})$ and
asymptotic behaviour $(v^{a})$. In technical terms, 
 no surface integrals remain when we calculate the  
  functional derivative of the action (2.19) with respect to the
 $\widetilde{E}$. By varying ${S}$ with respect to 
 $v^{\underline{a}}$ we obtain the constraint
\begin{equation}
\sum\limits_{\alpha} 
 \xi_{\alpha}^{\underline{a}} = 0\, .
\end{equation}

The choice (2.14) for  $\Lambda^{i}$ breaks asymptotically
the invariance with respect to local space translations (1.4). 
However, as  under general coordinate transformations the
functions $\widetilde{\Lambda}^{i}$ are scalars, the
changes $\delta \widetilde{\Lambda}^{i}$ under (1.4) of both the
singular and regular parts of $\widetilde{\Lambda}^{i}$ must
vanish and we obtain
\begin{equation}
\delta x^i = \delta a^{i} (t)\, ,
\end{equation}
where $x^i$ as well as $a^i$ transform as vectors under
rotations in tangent space.
Therefore as a residual symmetry we get translations, local in
time, and rigid rotations.

Let us calculate the corresponding generator of constant
translations (2.21). 
We find by using Noether's theorem  
\begin{eqnarray}
P^i & =& \sum\limits_{\alpha} p^i_{\alpha} + {2\over \lambda}
\int d^2 x  \, B^{\underline{a}} 
\widetilde{E}^{\underline{a}} _{i}
\cr\cr
& = & P^{i}_{\rm part} + P^{i}_{\rm field},
\end{eqnarray}
where the canonical particle momenta are obtained from (2.2),
(2.9) and (2.14) as
\begin{equation}
p^{j}_{\alpha} = \xi^{j}_{\alpha}
- {\lambda \over 4\pi} \sum\limits_{\beta  \atop
\beta \neq \alpha}
\left(
 \xi^{\underline{a}}_{\alpha}
 \cdot
  \xi^{\underline{a}}_{\beta}\right) 
  \partial_{j} \Phi ( \vec x_{\alpha \beta})
  \, ,
  \end{equation}
  where $\vec x_{\alpha \beta} = \vec x_{\alpha} - \vec x_{\beta}$. 
  
  Inserting the explicit expressions for $B^{\underline{a}}$ and
    $\widetilde{E}^{\underline{a}}_{i}$ into (2.22) we obtain
    \begin{equation}
    P^{i}_{\rm field} = 0 \, .
    \end{equation}
    Furthermore, from (2.23) we get
        \begin{equation}
    P^{i}_{\rm part} = \sum\limits_{\alpha} \xi^i_{\alpha}.
    \end{equation}

    The formula (2.25) describes the linear momentum of the
CM-motion for the N-particle system 
    which vanishes due to the 
    constraint (2.20)\footnote{Analogous situation has been observed in (2+
1)-dimensional gravity [13].}.
The same result has been obtained by one of the present authors
(PCS) in the one-dimensional case  [ 8], [ 9 ].

Next we pass to the Hamiltonian formulation. Applying the
Legendre transformation to the Lagrangian (2.19) and using the constraint

\begin{equation}
B^{\underline{a}} = \epsilon_{ij} \partial_j \, h^{\underline{a}}_{i} =
- {\lambda \over 2} 
\sum\limits_{\alpha} 
 \xi_{\alpha}^{\underline{a}}\,\delta^{(2)}
(x -x_{\alpha} )
\end{equation}
given by the components $\mu = j$, 
$\nu = i$ of (2.5) we find
 the N-particle Hamiltonian to be given by
\begin{equation}
H^{\rm (N)} = {1\over 2}
\sum\limits_{\alpha} 
 \xi_{\alpha}^{\underline{a}}\,
  \xi_{\alpha}^{\underline{a}}\,
  +
 v^{\underline{a}} \cdot
 \sum\limits_{\alpha} 
 \xi_{\alpha}^{\underline{a}}\, .
\end{equation}

\subsection{Hamiltonian Formulation for the Two--Body Problem}

Let us consider now more explicitly the  $N=2$ case.

We define:

\begin{equation}
\xi\sp{\underline{a}}:= {1\over 2} \left(
\xi_1\sp{\underline{a}}\, -\,\xi_2\sp{\underline{a}}\right) \, ,
\qquad 
x\sp{\underline{a}}:=\, x_1\sp{\underline{a}}\,
-\,x_2\sp{\underline{a}}\, ,
\qquad 
p\sp{\underline{a}}:=\,{1\over2}( p_1\sp{\underline{a}}
\, -\,p_2\sp{\underline{a}})\, .
\end{equation}
Then using the constraint (2.20) we get from (2.27)
\begin{equation}
H^{(2)}\,=\, \xi\sp{\underline{a}}\cdot \xi\sp{\underline{a}}
\label{2.29}
\end{equation}
and
\begin{equation}
p_i\,=\, 
\xi_i\,+\,{\lambda\over 4\pi}
(\xi\sp{\underline{a}}\cdot \xi\sp{\underline{a}})
\partial_i\,\Phi(\vec{x})
\, .
\label{eq}
\end{equation}

 The
Hamiltonian equations take the form

\setcounter{equation}{0}
\renewcommand{\theequation}{\thesection.31\alph{equation}}
\begin{equation}
\dot{x}\sp{i}\,=\, {\partial H\over \partial p\sp{i}}\,=\,{2}
\left(\xi\sp{\underline{a}}\cdot {\partial \xi\sp{\underline{a}}\over 
\partial p\sp{i}}\right),
\end{equation}

\begin{equation}
\dot{p}_{i}\,=\,- \,  {\partial H\over \partial x_{i}}\,= - \,{2}
\left(\xi\sp{{a}}\cdot {\partial \xi\sp{{a}}\over 
\partial x\sp{i}}\right).
\end{equation}

Using 
  (\ref{eq})  we have
\renewcommand{\theequation}{\thesection.32\alph{equation}}
\setcounter{equation}{0}
\begin{equation}
\xi\sp{\underline{a}}\cdot {\partial \xi\sp{\underline{a}}\over 
\partial p\sp{i}} 
\,=\,{\xi_i\over 
1+ \displaystyle{\lambda\over 2\pi}
(\xi\sp{\underline{a}}\cdot \partial_a\Phi)},
\end{equation}

\begin{equation}
\xi\sp{{a}}\cdot {\partial \xi\sp{{a}}\over 
\partial x\sp{j}}
\,=\,\ \, 
-\, {\lambda \over 4\pi} \
{\left(\xi_a  \xi^a \right) \xi^i\partial_i \partial_j \Phi \over 
1+  \displaystyle{\lambda\over 2\pi}(\xi\sp{{a}}\cdot \partial_a\Phi)}.
\end{equation}
\renewcommand{\theequation}{\thesection.\arabic{equation}}
\setcounter{equation}{32}

Taking the time derivative of 
 (\ref{eq})  and using (2.31--32) we get
\begin{equation}
\dot{\xi}\sp{{a}}_{i} + {\lambda \over 2\pi} \partial_{i}
\Phi(x) 
\xi_{j}\dot{\xi}_{j} = 0 \, .
\end{equation}
However, instead of the canonical variables ($x_i, p_i$) we can
use the variables ($x_i, \xi_i$). Then the Lagrangian 
obtained from the Hamiltonian 
  (\ref{2.29})  would have had the form:
\begin{eqnarray}
L & = &  p_l (\xi_i , x_l )\cdot \dot{x}_l - H 
\cr
& = & \left( \xi_l + {\lambda \over 4\pi} \xi^2
\partial_l \Phi(x)\right)\dot{x}_{l} -  \xi^2\, .
\end{eqnarray}
The variation with respect to $\xi_i$ is given by the expression
\begin{equation}
\xi_i = \displaystyle { { 1\over 2} \dot{x}_{i} \over 1 
- \displaystyle{\lambda\over 4\pi} \left(
\dot{x}_{j}\partial_j \Phi \right) },
\end{equation}
which is equivalent to the Hamiltonian eq. (2.31a) with the
insertion of (2.32a). Inserting (2.35) in (2.34) we get
\begin{equation}
L = {1\over 4} \  \displaystyle{\dot{x}^{2}_{l} \over
  \displaystyle 
  1 - \displaystyle{\lambda\over 4\pi} \left( \dot{x}_{l} \partial_l \Phi
\right) } \, .
\end{equation}

In particular if we consider
\begin{eqnarray}
\det\left( {\partial^2 L \over \partial \dot{x}_i \partial \dot{x}_j }\right)
&= &
{1\over 4} \left(1 - {\lambda\over 4\pi} \left( \dot{x}_{l} \partial_l \Phi
\right) \right)^{-4}
\cr
&= &
{1\over 4} \left(1 + {\lambda\over 2\pi} 
 \xi_{l} \partial_l \Phi
 \right)^{+4}\, ,
 \end{eqnarray}
we see that when the 
 velocities are expressible in terms of the canonical variables
then
(2.33) gives us
\begin{equation}
\dot{\xi}_l = 0\, .
\end{equation}
Using the Hamiltonian 
 (\ref{2.29})  we get for the pair of
noncanonical variables the Hamilton equations (2.31a), as
$\dot{x}_l = \{ x_l , H \}$ and (2.38) as 
$\dot{\xi}_l = \{ \xi_l , H \}$, if we assume the following
nonstandard symplectic structure:

\setcounter{equation}{0}
\renewcommand{\theequation}{\thesection.39\alph{equation}}
\begin{equation}
\left\{ x^i , x^j \right\} = \left\{ \xi_i, \xi_j \right\} = 0,
\end{equation}

\begin{equation}
  \left\{ x^i , \xi_j  \right\}= \delta^{i}_{\ j}
- {{\lambda\over 2\pi} \xi^i \partial_j \Phi \over
1 +  \displaystyle
{\lambda \over 2\pi } \left( \xi^a \partial_a \Phi \right) }.
\end{equation} 

It is easy to check  that the Poisson brackets (2.39a--b) satisfy
the Jacobi identity.
For our system we have two conserved angular momenta.
If we define (in $D=2$ $\vec{a} \wedge \vec{b} = \epsilon_{ij}a^i
b^j$)
\begin{equation}
l:= \,\vec{x}\,\wedge {\vec{p}}\,
\,=\,
\vec{x}\,\wedge \vec{\xi}\,
+{\lambda\over 4\pi} H
\label{ang}
\end{equation}
we find that  $l$ is conserved as well as
 \begin{equation}
 \bar l:=\vec{x}\,\wedge \vec{\xi}\,
 \end{equation}
 because 
\begin{equation}
{d\over dt}\bar l\,=\,\dot{\vec{x}}\,\wedge 
\vec{\xi}\,
\,=\,0.
\end{equation}
However, as
\begin{equation}
\xi\sp{\underline{a}}\cdot \partial_a\,\Phi\,
=\,- \vec{\xi}\,\wedge
\vec{\nabla}\,\ln|x\sp{\underline{c}}|\,=\,{\bar l\over r\sp2},
\end{equation}
where $r:=|x\sp{\underline{a}}|$,  we see that (2.35) can be rewritten as:
\begin{equation}
\dot x\sp{\underline{a}}\,=\,{ 2\xi\sp{\underline{a}}\over
1+{\lambda\bar l\over 2\pi r\sp2}}.
\end{equation}

\renewcommand{\theequation}{\thesection.\arabic{equation}}
\setcounter{equation}{39}
Note that if $x\sp{\underline{a}}(0)$ is parallel to $\xi\sp{\underline{a}}$
we have $\bar l=0$ and, in consequence, a free motion on a line.

 For $\lambda \overline{l} < 0$ it is convenient to
introduce the quantity 
 \begin{equation}
 r^{2}_{0} : =  { | \lambda \overline{l} |  \over 2 \pi }
 \end{equation}
as then, from the previous results, we see that the relative two--body
motion separates in this case into a motion within the interior
region given by
\renewcommand{\theequation}{2.41\alph{equation}}
\setcounter{equation}{0}
\begin{equation}
r < r _0
\end{equation}
and a motion within the exterior region given by

\begin{equation}
r > r _0.
\end{equation}

\renewcommand{\theequation}{\thesection.\arabic{equation}}
\setcounter{equation}{41}

\section{New Planar Dynamics: Quantum--Mechanical Two--Body Problem}
The relation
(\ref{eq}), 
 after the substitution of  
 (\ref{2.29}),
takes the form ($H\equiv H^{(2)}$):

\begin{equation}
\vec{\xi}\,=\,\vec{p}\,-\,{\lambda\over 4\pi} \, H
\,\vec{\bigtriangledown} \, \Phi.
\end{equation}

\setcounter{equation}{0}
\renewcommand{\theequation}{\thesection.2\alph{equation}}

Squaring it and using again (2.29) we obtain
\begin{equation}
H = \vec{p}\, ^{2} - {l^2 \over r^2 } +
 { \overline{l}\, ^2 \over r^2},
\label{eq4}
\end{equation}
where we have used the definition  (2.39d) of $\overline{l}$.   
 
 It should be stressed that
 \begin{equation}
 \overline{l} = l - {\lambda \over 4 \pi } H
 \end{equation}
 i.e. (3.2a) gives us a quadratic equation for $H$.

\setcounter{equation}{2}
\renewcommand{\theequation}{\thesection.\arabic{equation}}

We quantize the problem by considering a Schr\"odinger-like equation 
\begin{equation}
i\hbar {\partial \psi(\vec{x},t) \over 
\partial t} =    \hat{H} \, \psi (\vec{x},t) =
\left[
\hat{\vec{p}}\, \sp2- {l^2  \over r^2} +
{1\over  r^2 } \, \overline{l}\, ^{2} \right]
\Psi(\vec{x},t)
\label{eq6}
\end{equation}
in which the operators $\hat H$ and $\hat{\vec{p}}$ 
  are defined  by  the usual
quantization rules
\begin{equation}
\hat H\,= \, i\hbar {\partial\over \partial t},\qquad
 \hat{p}_{l}\,= \,
{\hbar\over i}\partial_l \, .
\end{equation}

We see that the equation (3.3) describes a nonstandard form of 
a time dependent  Schr\"{o}dinger equation, with its right hand side
  containing  both the first and second time derivatives (entering 
through $\overline{l}$).

For the stationary case, {\it i.e.}
 when $\Psi(\vec{x},t)=\Psi_E(\vec{x})e\sp{{iEt\over \hbar}}$
we can use the angular-momentum basis and put
\begin{equation}
\Psi_{E,m}\,=\,f_{E,m}(r ) \, e\sp{im\varphi}
\end{equation}
where $m$ is an integer, and find that $f_{E,m}$ satisfies a nonstandard
 time independent  Schr\"odinger equation
\begin{equation}
\left[-\hbar\sp2\left(\partial_r\sp2 
+ {1\over r}\partial_r 
-{\bar{m}\sp2\over r\sp2}\right)-E\right]f_{E,m}(r) =0,
\label{sch}
\end{equation}
where we have defined
\begin{equation}
\hbar \bar{m} :\,=\,\hbar m-{\lambda\over 4\pi} E
\label{bar}
\end{equation}
{\it i.e.} $\hbar \bar{m}$ is an eigenvalue of ${\overline{l}}$.

A characteristic feature of the Schr\"odinger equation (3.6) is the 
appearance
of the noncanonical angular momentum $\hbar\bar{m}$ with $\bar{m}$ not 
being an integer.  
{\it i.e.} our two-body system carries a fractional orbital angular
momentum (see the discussion in [10] or [14]). It can be shown
that $J= {\overline{l}}$ is equal to the total
angular momentum of the particle + field system with a
nonvanishing angular momentum of the fields [15]. 

In the following we discuss only the most interesting case of
$\lambda \overline{l} < 0$.

Now the appropriate boundary conditions correspond to the
requirement that $f_{E,m}(r) $ is nonzero in either the interior
region ($r < r_0$) or in the exterior region ($r > r_0$). Thus
our boundary condition is
\begin{equation}
f_{E,m}(r_0) = 0
\end{equation}
The general solution of (3.6) is given by

\begin{equation}
f_{E,m}(r) = 
 Z_{ \overline{m}} \left( {\sqrt{E} \over \hbar} \, r \right)
\, ,
\end{equation}
where $Z_{\overline{m}}$ is an appropriate Bessel function of
order $\overline{m}$ (or a superposition of such functions).

\subsection*{Interior solutions ($r < r_0 $)}

The only Bessel functions 
 belonging to a  self-adjoint differential operator in (3.6)
 are those of 
the first kind with $\bar{m}\ge 0$.
 Therefore a decomposition of space into interior/exterior regions
 appear only in the case of $\lambda < 0$.

Then the possible eigenvalues $E_n(m)$ are determined by
\begin{equation}
J_{\bar{m}}\left[{\sqrt{E}\over \hbar}
\left( {\hbar\vert \lambda\vert \bar{m}\over 2\pi}\right)
\sp{1\over 2}\right]\,=\,0
\label{result}
\end{equation}
with $\bar{m}$ given by (3.7).

  To simplify (3.10) we define
\begin{equation}
\epsilon\,=\,{\vert \lambda\vert E\over 2\pi \hbar}.
\end{equation}
Then (3.10) takes the form
\begin{equation}
J_{\bar{m}}(\bar{m}\sp{1\over 2}\epsilon\sp{1\over2})\,=\,0.
\end{equation}

As $J_{\bar{m}}$, for fixed $\bar{m}>0$, has
no imaginary zeros (see e.g. [16]) we have to 
 consider
 an infinite number of simple positive
 zeros, which we denote by $y_n(\bar{m})$, $n=1,2..$ 
arranged in ascending order of magnitude. Then 
we see that 
due to (3.7), the eigenvalues
 $\epsilon_n(m)$ we are looking for, are the 
 positive fixed points of the equation
\begin{equation}
\epsilon\,=\,f_n(m+{1\over 2}\epsilon),
\label{neweq}
\end{equation}
subject to the conditions $m+{1\over 2}\epsilon_n (m)>0$ and 
 $y_{n}(\bar{m}) > 0$, 
 where we have defined 
\begin{equation}
f_n(\bar{m}):=\,{1\over \bar{m}}y_n\sp2(\bar{m}).
\end{equation}
The solutions of (3.13) can be obtained  only numerically. 

We insert (3.14) into (3.13) and after the use of approximate
formulae for the zeros of Bessel functions of first kind [16] we
conclude that the low-lying energy levels are given by small values
of $n$. 
In this case all values of $m\in \zetZ$ contribute to the energy
spectrum ($E > 0$) but the positivity of $\bar{m} = m+{1\over 2}
\epsilon_n (m) > 0$ for higher $n$ implies that some finite
range of values of $m$ has to be excluded. Qualitatively one can
say that for every $n$ there appears however an infinite 
 tower 
 of energy values $\epsilon_n (m)$.

Only
in the case of large $m$ the analytical asymptotic results are 
available. Let us
consider the choice 
 $n=1$.
Using the asymptotic behaviour ([16])

\begin{equation}
y_1 (\overline{m}) = \overline{m}
+ 1.855757\overline{m}^{1\over 3} 
+ O(\overline{m}\, ^{- {1\over 3}}),
\end{equation}
valid for large $\overline{m}$, we obtain
\begin{equation}
\epsilon_1 ({m}) = 2{m}
+ 9.35243{m}^{1\over 3} + O({m}^{- {1\over 3}})\ .
\end{equation}
valid for large positive $m$.

\subsection*{Exterior solutions $(r>r_0)$}

First of all, let us note that there are no bound states solutions 
of (3.6) for $r>r_0$ as the only square integrable 
Bessel functions in $[r_0,\infty)$ are the
 modified
Bessel functions of the third kind, 
which have neither positive nor pure imaginary zeros (see {\it e.g.} [16]).

Scattering solutions are given by 
a superposition of Bessel functions of the first and second kind

\begin{equation}
f_{E,m}(r )\,=\,A_m(E)\,J_{\bar{m}}\left({\sqrt{E}\over \hbar}r\right)\,+\,
B_m(E)\,Y_{\bar{m}}\left({\sqrt{E}\over \hbar}r\right)
\end{equation}
with the ratio ${A_m\over B_m}$ determined by the boundary condition 
(3.8).
These solutions describe scattering
  on an obstruction of radius $r_0$, which is
dynamically determined.

\section{Final Remarks}

Gauging of translations for nonrelativistic point particles in
a $D=2$ dimensional space coupled to a translational 
Chern--Simons action leads to a
new and interesting $D=(2+1)$ particle dynamics. In particular, we
have shown that for negative values of the product $
\lambda \overline{l}$ of the coupling strength $\lambda$ and the
fractional orbital angular momentum $ \overline{l}$
the two--body motions separate into interior and exterior
motions. In the quantum mechanical case we have obtained a
 new type of
nonlinear Schr\"{o}dinger equation, with a second order time
derivative, which leads, in the stationary case, to the nonlinear energy
eigenvalue problem. The interior solutions are described by a
infinite tower of bound states for each fixed 
$n$.  They describe confinement within a dynamically determined
compact space region (see (2.41a)) (geometric bag). We should
mention that analogous  results were obtained by
one of the present authors (PCS) in the one--dimensional case [9].
It is clear that a possible physical relevance of these results, which recall
the features of confinement  in one and two
space dimensions, should be investigated further.

\end{document}